\documentclass[aps,showpacs,showkeys,preprintnumbers,amsmath,amssymb]{revtex4}

\usepackage{amsmath,amssymb,amsthm}
\usepackage{color}
\usepackage{epsfig}
\usepackage{multirow}
\usepackage{graphicx}
\usepackage{subfigure}
\usepackage{times}
\usepackage{chngpage}
\usepackage{caption}
\usepackage{rotating}
\usepackage{tabularx}
\usepackage{booktabs}
\usepackage[lofdepth,lotdepth]{subfig}
\usepackage{dcolumn}
\usepackage{mathrsfs}
\usepackage{bm}
\usepackage{epsfig,float,graphics}

\begin{document}

\title{Strong gravitational lensing for the charged black holes with scalar hair}
\author{Ruanjing Zhang, Jiliang Jing\footnote{Corresponding author, Email: jljing@hunn.edu.cn}}
\affiliation{Department of
Physics, Key Laboratory of Low Dimensional Quantum Structures and
Quantum Control of Ministry of Education, and Synergetic Innovation
Center for Quantum Effects and Applications, Hunan Normal
University, Changsha, Hunan 410081, P. R. China
}

\begin{abstract}
The strong gravitational lensing for charged black holes with scalar hair in Einstein-Maxwell-Dilaton theory are studied.
We find, with the increase of scalar hair, that the radius of the photon sphere, minimum impact parameter, angular image position and relative magnitude increase, while the deflection angle and angular image separation decrease. Our results can be reduced to those of the Schwarzschild black hole in two cases, one of them is that the scalar hair disappears, the other is that the coupling constants take  particular values with arbitrary scalar hair.
\end{abstract}


\maketitle

\section{Introduction}

The presence of a massive body produces the deflection of light passing close to the object according to the theory of general relativity;
the corresponding effects are called as gravitational lensing, and the object causing a detectable deflection acts as gravitational lens  \cite{Einstein}. What's more, this deflection of light was first observed in 1919 by Dyson, Eddington, and Davidson \cite{Dyson}. After the pioneering strong gravitational lensing in Q0957+561 \cite{Walsh} was discovered in 1979, gravitational lensing developed into an important astrophysical tool to extract information about distant stars which are too dim to be observed, similar to a natural and large telescope.
When an object with a photon sphere is situated between a source and an observer, there are two infinite sets of images called relativistic images, produced by light passing close to the photon sphere, which undergoes a large deflection.
It is shown that these relativistic images carry much valuable information about the central celestial objects and could provide the profound verification of alternative theories of gravity \cite{Vir,Fritt,Bozza2,Eirc1,whisk,Gyulchev,Bhad1,TSa1,AnAv,gr1,Kraniotis,JH,Bozza4}.
Therefore, gravitational lensing is regarded as a powerful indicator of the physical nature of the central celestial object. So, we need a systematic approach to calculate the deflection angle and the feature of relativistic images. Darwin \cite{Darwin} calculated the deflection angle by using the strong deflection limit (consisting of a logarithmic approximation)  for the Schwarzschild spacetime. And this method allows for calculating the position, magnification of the relativistic images. It was rediscovered several times \cite{Bozza3}, then extended to the Reissner-Nordstr\"{o}m metric \cite{Eirc}, and to any spherically symmetric objects with a photon sphere \cite{Bozza}. In recent years, many works \cite{schen,zhang} have been done basing on this method.

The standard ``no-hair theorem" \cite{Ruffini} states that a black hole is completely specified by the mass, charge, and angular momentum. However, during the recent years, much attention was devoted to gravity theories supplying by scalar field, and many examples of scalar hairy black holes \cite{Nadalini,Anabalon,Herdeiro,Martinez} have been obtained. There are several reasons for this. To begin with, as one kind of the fundamental and effective fields, scalar field is well motivated by standard-model particle physics. In addition, we analyze different field contents which can be treated as a means of checking the ``no-hair theorem" and exploring the structure of black holes. Scalar field is often considered by physicists as one of the simplest types of ``matter". At last, the presence of the scalar field leads to different black hole spacetimes, which may engender some new phenomena. We hope these deviations could be detected in astrophysical observations. What's more, the fundamental scalar field does exist in nature by discovering a scalar particle at the Large Hadron Collider at CERN \cite{Aad,Chatrchyan}, which has been identified as the standard-model Higgs boson since 2012. Therefore, to study strong gravitational lensing and time delay for black holes with scalar hair has great significance.

This paper is arranged as follows: In Sec. \ref{section 2}, we study the physical properties of strong gravitational lensing around the charged black holes with scalar hair and probe the effects of the scalar hair on the event horizon, the radius of the photon sphere, the minimum impact parameter, and the deflection angle. In Sec. \ref{section 3}, we suppose that the gravitational field of the supermassive black hole at the centre of our Galaxy can be described by this metric, and then obtain the numerical results for the main observables in strong gravitational lensing, such as the angular image position, the angular image separation, and the relative magnitude of relativistic images. Finally, we will include our conclusions in the last section.

\section{Deflection angle for the charged black holes with scalar hair}\label{section 2}

Supergravities have provided a variety of fundamental matter fields that we can study their interactions with gravity \cite{Sagnotti,Duff,Erler,Faedo,Dall}, one of them is dilatonic scalar \cite{Gibbons}. If the dilaton $\varphi$ couples to an $n$-form field strength $F_{n}=dA_{n-1}$ through $Z(\varphi)$, the general class of Lagrangian is given by \cite{Fan}
\begin{equation}\label{L}
e^{-1}\mathcal{L}=R-\frac{1}{2}(\partial\varphi)^{2}-\frac{1}{2n!}Z(\varphi)F^{2}_{n},
\end{equation}
where $e=\sqrt{-det(g_{\mu\nu})}$. It is not hard to find that we can get the usual Reissner-Nordstr\"{o}m black hole decoupled  with the dilaton if $Z(\varphi)$ in Eq. (\ref{L}) has a stationary point. And the uniqueness theorem is broken if we can construct a further different black hole with the same mass and charge, but non-vanishing dilaton. Now, we suppose $Z$ as
\begin{equation}\label{Z}
Z^{-1}=e^{a_{1}\varphi}\cos^{2}\omega +e^{a_{2}\varphi}\sin^{2}\omega ,
\end{equation}
with
\begin{equation}\label{a10}
a_{1}a_{2}=-\frac{2(n-1)(D-n-1)}{D-2},
\end{equation}
where $a_1$ and $a_2$ are the dilaton coupling constants, and $\omega$ is another coupling constant.  The function $Z$ becomes an exponential function of $\varphi$ for $\omega=0$ or $\omega=\frac{\pi}{2}$.

In this paper,   we focus our attention  on the Einstein-Maxwell-Dilaton theory in four dimensions, corresponding to $D = 4$ and $n = 2$, in which the dilaton $\varphi$ coupling to the Maxwell field $A$ is not the usual single exponential function, but one with a stationary point. The condition for $(a_{1}, a_{2})$ in Eq. (\ref{a10}) becomes $a_{1}a_{2}=-1$, and then the Lagrangian can be rewritten as
\begin{equation}\label{L1}
e^{-1}\mathcal{L}=R-\frac{1}{2}(\partial\varphi)^{2}-\frac{1}{4}ZF^{2},
\end{equation}
where $F=dA$. The constants  $a_{1}, \ a_{2}$  can be expressed as
 \begin{equation}\label{a1}
a_{1}=\sqrt{\frac{1-\mu}{1+\mu}},~~~~a_{2}=-\sqrt{\frac{1+\mu}{1-\mu}},
\end{equation}
where $\mu$ is a dimensionless constant with the range of $\mu\in(-1,1)$.
The dilaton coupling function $Z$ is thus given by
\begin{equation}\label{Z1}
Z^{-1}=e^{\sqrt{\frac{1-\mu}{1+\mu}}\varphi}\cos^{2}\omega +e^{-\sqrt{\frac{1+\mu}{1-\mu}}\varphi}\sin^{2}\omega .
\end{equation}
Then, the Lagrangian (\ref{L1}) admits the charged black holes with scalar hair as  \cite{Fan}
\begin{equation}\label{metric1}
ds^{2}=-f(r)dt^{2}+\frac{dr^{2}}{f(r)}+r^{1+\mu}(r+S)^{1-\mu}(d\theta^{2}+\sin^{2}\theta d\phi^{2}),
\end{equation}
where
\begin{equation}\label{metric2}
f(r)=\left(1+\frac{S}{r}\right)^{\mu}\left[1+\frac{Q^{2}\cos^{2}\omega }{2rS(1+\mu)}-\frac{Q^{2}\sin^{2}\omega }{2S(1-\mu)(r+S)}\right].
\end{equation}
The solution involves two integration constants, one is constant $Q$ which parameterizes the electric charge, the other is $S$ that associates with dilaton $\varphi$ by
\begin{equation}\label{phi}
e^{\frac{\varphi}{\sqrt{1-\mu^{2}}}}=1+\frac{S}{r}.
\end{equation}
Since the black hole has scalar hair with varying $\varphi$,  $S$ parameterizes the scalar hair.

What important is that this solution will return to the Schwarzschild black hole when $S\rightarrow0$ or $\omega=\frac{\pi}{2}$ with $\mu\rightarrow1$. Hence, this limit can be used to test our results in the following study. After that, the ADM mass, electric charge, and Maxwell field $A$ are given by  \cite{Fan}
\begin{equation}\label{m}
M=\frac{Q^{2}\sin^{2}\omega}{4(1-\mu)S}-\frac{Q^{2}\cos^{2}\omega}{4(1+\mu)S}-\frac{1}{2}\mu S,~~~~~Q_{e}=\frac{1}{4}Q,~~~~~A=\frac{Q(r+S\cos^{2}\omega)}{r(r+S)}.
\end{equation}
It is useful for the calculation to follow the scaling symmetries in the forms
\begin{equation}\label{scaling}
\frac{r}{2M}\rightarrow r, ~~\frac{S}{2M}\rightarrow S, ~~\frac{Q}{2M}\rightarrow Q, ~~\frac{t}{2M}\rightarrow t, ~~\frac{C(r)}{(2M)^{2}}\rightarrow C(r).
\end{equation}
After taking the scaling symmetries, the solution (\ref{metric1}) still takes the same form as above. Since it is meaningful to study the effects of scalar hair on the strong gravitational lensing, we can use $\mu$, $\omega$, $S$ to show $Q^{2}$ as

\begin{equation}\label{Q}
Q^{2}=\frac{2S(1-\mu)(1+\mu)(1+S\mu)}{(\mu-1)\cos^{2}\omega+(\mu+1)\sin^{2}\omega}.
\end{equation}
Then, we have to take $\omega\in[\frac{\pi}{4},\frac{\pi}{2}]$ with $\mu=0$ or $\mu\in(-1,1)$  with $\omega=\frac{\pi}{2}$ to ensure $Q^{2}>0$.

\begin{figure}[htbp]
\begin{center}
\includegraphics[scale=1]{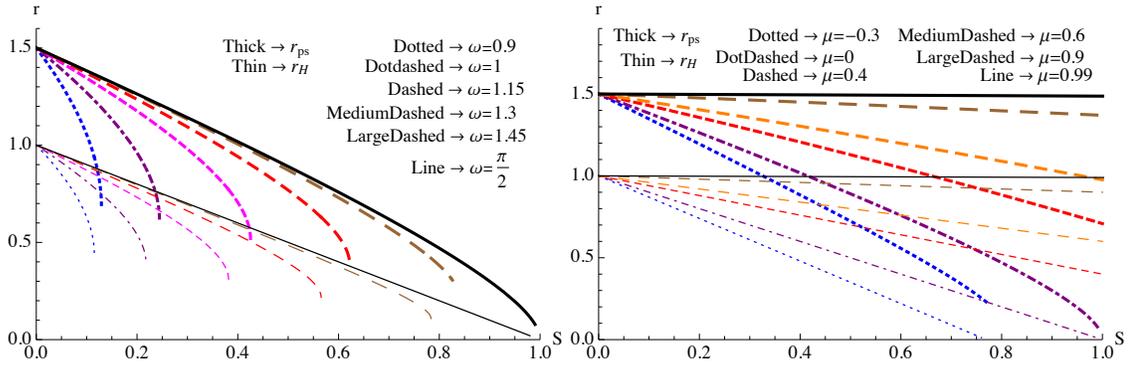}
\end{center}
\caption{Radius of the photon sphere (thick line) and event horizon (thin line) change with scalar hair $S$ for $\mu=0$ (left) and $\omega=\frac{\pi}{2}$ (right).}
\label{rps00}
\end{figure}

Now, let us study the physical properties of strong gravitational lensing by the charged black holes with scalar hair. We choose the equatorial plane ($\theta=\frac{\pi}{2}$) which means that both the observer and the source lie in the equatorial plane, and the whole trajectory of the photon is limited on the same plane. Then the metric (\ref{metric1}) can be expressed as
\begin{equation}\label{metric3}
ds^2=-A(r)dt^2+B(r)dr^2+C(r)d\phi^2,
\end{equation}
with
\begin{equation}\label{metric4}
A(r)=f(r), ~~B(r)=\frac{1}{f(r)}, ~~C(r)=r^{1+\mu}(r+S)^{1-\mu}.
\end{equation}
In the spherically symmetric case, the equation of the photon sphere reads
\begin{equation}\label{u1}
\frac{C'(r)}{C(r)}=\frac{A'(r)}{A(r)},
\end{equation}
where the prime represents the derivative with respect to $r$. For the charged black holes with scalar hair, the equation of the photon sphere takes the form
\begin{eqnarray}\label{u2}
&& 8Sr^{3}(\mu^{2}-1)+r^{2}[4S^{2}(2\mu+3)(\mu^{2}-1)+6Q^{2}(\mu-\cos2\omega)]\nonumber \\
&& +2Sr[2S^{2}(2\mu^{3}+\mu^{2}-2\mu-1)+Q^{2}(2\mu^{2}+3\mu-2-3\cos2\omega)]\nonumber \\  && +2S^{2}Q^{2}(1+\cos2\omega )(\mu^{2}-1)=0.
\end{eqnarray}
Obviously, this equation has three roots because it is a cubic equation of $r$. We take the root tends to 1.5 when $S\rightarrow0$ as the radius of the photon sphere. In other words, we take the root which could return to the Schwarzschild black hole when $S\rightarrow0$. We present the variation of the radius of the photon sphere $r_{ps}$ and radius of the event horizon $r_{H}$ with the scalar hair $S$ for $\mu=0$ (varying $\omega$) and $\omega=\frac{\pi}{2}$ (varying $\mu$) in Fig. \ref{rps00}.  We can see that $r_{H}$ and $r_{ps}$ are both decrease with the increase of scalar hair for either $\mu=0$ or $\omega=\frac{\pi}{2}$. And we can also see $r_{ps}$ always bigger than $r_{H}$ for given $\mu$ and $\omega$, this is in accordance with our usual perception. This black hole can recover to the Schwarzschild black hole \cite{Bozza} ($r_{H}=1$, $r_{ps}=1.5$) in two cases, one is $S\rightarrow0$ for arbitrary $\omega$ and $\mu$, another is $\omega=\frac{\pi}{2}$ and $\mu\rightarrow1$ for arbitrary scalar hair $S$. From Fig. \ref{rps00}, we can see that each line of $r_{H}$ and $r_{ps}$ intersects when $S\rightarrow0$, and the black line in the right graph is basically a straight line, these are both performances of recovering to the results of the Schwarzschild case.

The exact deflection angle $\alpha$ for a photon coming from infinity relates to the closest approach distance $r_{0}$ can  be expressed as \cite{Weinberg}
\begin{equation}\label{angle1}
\alpha(r_{0})=I(r_{0})-\pi,
\end{equation}
with
\begin{equation}\label{angle2}
I(r_{0})=2\int^{\infty}_{r_{0}}\frac{\sqrt{B(r)}dr}{\sqrt{C(r)}\sqrt{\frac{C(r)A(r_{0})}
{C(r_{0})A(r)}}-1}.
\end{equation}
The deflection angle is a monotonic decreasing function of $r_{0}$.
For a special value of $r_{0}$, the deflection angle will become $2\pi$, that is to say, the light ray will makes a complete loop around the compact object before reaching the observer, which results in two infinite sets of relativistic images, one is on the same side, and the other is on the opposite side of the source.
Furthermore, the deflection angle diverges when $r_{0}$ approaches to the radius of the photon sphere  $r_{ps}$, which means that the photon is captured.

We are now in position to calculate the case of a photon passing close to the photon sphere, by using the evaluation method for the integral (\ref{angle2}) proposed by Bozza \cite{Bozza}. Then, it is useful to define a new variable
\begin{equation}\label{variable}
z=1-\frac{r_{0}}{r},
\end{equation}
and we will obtain
\begin{equation}\label{angle3}
I(r_{0})=\int^{1}_{0}R(z,r_{0})f(z,r_{0})dz,
\end{equation}
with
\begin{equation}\label{R}
R(z,r_{0})=\frac{2r^{2}\sqrt{A(r)B(r)C(r_{0})}}{r_{0}C(r)},\nonumber
\end{equation}
\begin{equation}\label{f}
f(z,r_{0})=\frac{1}{\sqrt{A(r_{0})-\frac{A(r)C(r_{0})}{C(r)}}},
\end{equation}
where $R(z,r_{0})$ is the regular term, and $f(z,r_{0})$ is the divergent term which diverges for $z\rightarrow0$---i.e., the photon approaches the photon sphere. So we can split the integral (\ref{angle3}) as a sum of two parts
 \begin{eqnarray}\label{IDR}
I_{D}(r_{0})&=&\int^{1}_{0}R(0,r_{ps})f_{0}(z,r_{0})dz, \\
I_{R}(r_{0})&=&\int^{1}_{0}[R(z,r_{0})f(z,r_{0})-R(0,r_{ps})f_{0}(z,r_{0})]dz,
\end{eqnarray}
where $I_{D}(r_{0})$ and $I_{R}(r_{0})$ denote the divergent and regular parts in the integral (\ref{angle3}), respectively. To find the order of divergence of the integrand, we take a Taylor expansion of the argument of the square root in $f(z,r_{0})$ to the second order in $z$; then we get
\begin{equation}\label{f0}
f_{0}(z,r_{0})=\frac{1}{\sqrt{p(r_{0})z+q(r_{0})z^{2}}},
\end{equation}
with
\begin{eqnarray}\label{pq}
p(r_{0})&=&\frac{r_{0}}{C(r_{0})}(A(r_{0})C'(r_{0})-A'(r_{0})C(r_{0})),\\\nonumber
q(r_{0})&=&\frac{r_{0}}{2C^{2}(r_{0})}(2r_{0}C(r_{0})C'(r_{0})A(r_{0})-2r_{0}C'^{2}(r_{0})A(r_{0})\\
&+&r_{0}C(r_{0})C''(r_{0})A(r_{0})-r_{0}C^{2}(r_{0})A''(r_{0})).
\end{eqnarray}
It is obviously that $p(r_{0})=0$ at $r_{0}=r_{ps}$ from Eqs. (\ref{u1}) and (\ref{pq}). So we have $f_{0}(z,r_{0})\sim\frac{1}{z}$ when $r_{0}$ is equal to the radius of the photon sphere $r_{ps}$, and then the term $I_{D}(r_{0})$ diverges logarithmically. Therefore, the deflection angle can be expanded in the form
\begin{equation}\label{angle4}
\alpha(\theta)=-\bar{a}\log(\frac{\theta D_{OL}}{u_{ps}}-1)+\bar{b}+o(u-u_{ps}),
\end{equation}
with
\begin{eqnarray}\label{ab}
\nonumber \bar{a}&=&\frac{R(0,r_{ps})}{2\sqrt{q(r_{ps})}},\\
\nonumber  \bar{b}&=&-\pi+b_{R}+\bar{a}\log\frac{r_{ps}^{2}[C''(r_{ps})A(r_{ps})-C(r_{ps})A''(r_{ps})]}{u_{ps}\sqrt{A^{3}(r_{ps})C(r_{ps})}},\\
\nonumber b_{R}&=&I_{R}(r_{ps}),\\
u_{ps}&=&{\sqrt\frac{C(r_{ps})}{A(r_{ps})}},
\end{eqnarray}
where the quantity  $D_{OL}$ is the distance between the observer and the gravitational lens; $\theta$ is the angular separation between the optical axis and the direction of image which satisfies $u=\theta D_{OL}$; $u_{ps}$ is the impact parameter $u$ evaluated at $r_{ps}$ which is called the minimum impact parameter; $\bar{a}$ and $\bar{b}$ are strong deflection limit coefficients which depend only on the metric function evaluated at $r_{ps}$. Making use of Eqs. (\ref{angle4}) and (\ref{ab}), we can study the properties of strong gravitational lensing by the charged black holes with scalar hair.

\begin{figure}[htbp]
\begin{center}
\includegraphics[scale=1]{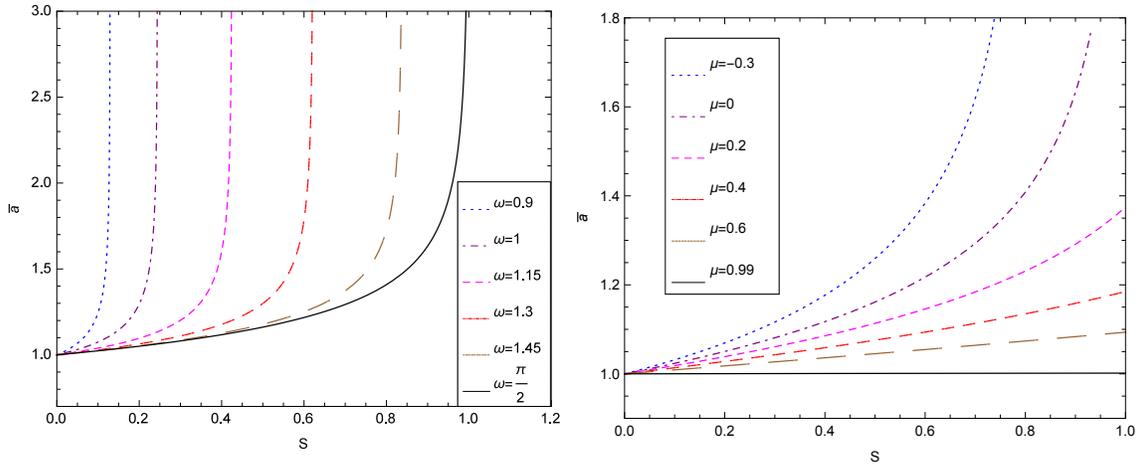}
\end{center}
\caption{Variation of the coefficient $\bar{a}$ with scalar hair $S$ for $\mu=0$ (left) and $\omega=\frac{\pi}{2}$ (right).}
\label{a}
\end{figure}

\begin{figure}[htbp]
\begin{center}
\includegraphics[scale=1]{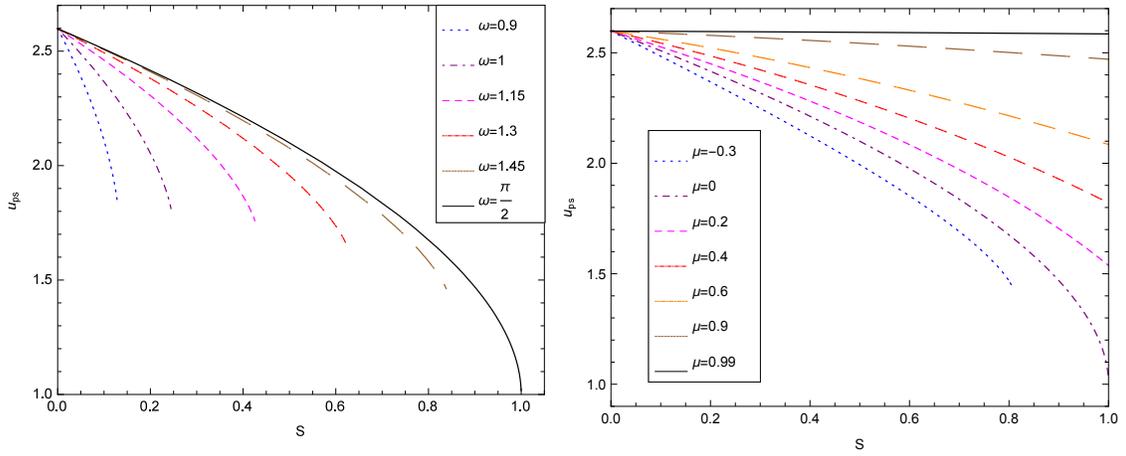}
\end{center}
\caption{Variation of the minimum impact parameter $u_{ps}$ with scalar hair $S$ for $\mu=0$ (left) and $\omega=\frac{\pi}{2}$ (right).}
\label{u}
\end{figure}

\begin{figure}[htbp]
\begin{center}
\includegraphics[scale=1]{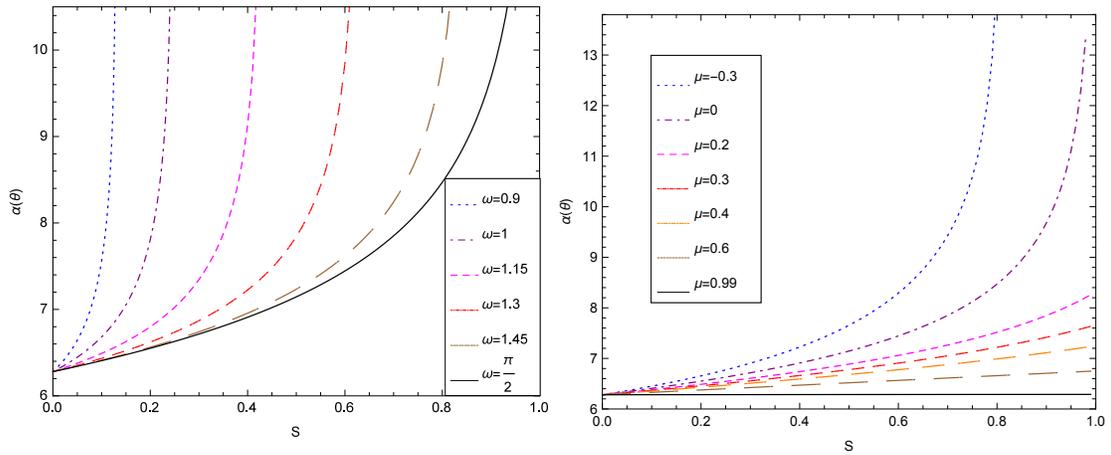}
\end{center}
\caption{The deflection angle evaluated at $u=u_{ps}+0.003$ as a function of scalar hair $S$ for $\mu=0$ (left) and $\omega=\frac{\pi}{2}$ (right).}
\label{aa}
\end{figure}

Now, we probe the properties of strong gravitational lensing by the charged black holes with scalar hair and mainly explore the effects of the scalar hair $S$ on the deflection angle. We  show, in Figs. \ref{a}-\ref{aa}, the variation of the coefficient $\bar{a}$, the minimum impact parameter $u_{ps}$, and the deflection angle $\alpha(\theta)$ with scalar hair $S$ for the change of $\omega$ when $\mu=0$, and for the change of $\mu$ when $\omega=\frac{\pi}{2}$, respectively. We can read from Fig. \ref{a} that the coefficient $\bar{a}$ always grows with the increase of scalar hair $S$  for either $\mu=0$ or $\omega=\frac{\pi}{2}$, but the growth rate decreases with the increase of $\mu$ or $\omega$. Furthermore, we get that the minimum impact parameter $u_{ps}$ decreases with the increase of scalar hair $S$ for either $\mu=0$ or $\omega=\frac{\pi}{2}$ in Fig. \ref{u}. We also plot the deflection angle $\alpha(\theta)$ evaluated at $u=u_{ps}+0.003$ in Fig. \ref{aa}. And then, we find that the deflection angle increases with the increase of scalar hair regardless of the varying $\mu$ and $\omega$, which tells us that the scalar hair enhances the effects of the black hole on the light. It is also shown that the deflection angle has the similar properties with the coefficient $\bar{a}$; this means that the deflection angle of the light ray is dominated by the logarithmic term in strong gravitational lensing.  There is one thing that we can't ignore is every line of $\bar{a}$, $u_{ps}$, and $\alpha(\theta)$ intersects at $S\rightarrow0$ for arbitrary $\mu$ and $\omega$, which implies that they recover to the results of the standard Schwarzschild case \cite{Bozza}; i.e., $\bar{a}=1$, $u_{ps}=2.598$, and $\alpha(\theta)=6.28$. We should also note that  the black lines in each graph on the right side of Figs. \ref{a}-\ref{aa} are almost straight lines,  it means that our results  recover to the results of the Schwarzschild case again for $\omega=\frac{\pi}{2}$ and $\mu\rightarrow1$.

\section{Observables in strong gravitational lensing}\label{section 3}

In this part, we calculate the observables in strong gravitational lensing by the charged black holes with scalar hair, including the angular image position $\theta_{\infty}$, the angular image separation $s$, and the relative magnitude $r_{m}$.
Let us start with the lens equation \cite{Bozza}
\begin{equation}\label{gamma}
\beta=\theta-\frac{D_{LS}}{D_{OS}}\triangle\alpha_{n},
\end{equation}
where $\beta$ is the angle between the direction of the source and the optical axis, called the angular source position. $D_{LS}$ is the distance between the lens and the source; $D_{OS}$ is the distance between the observer and the source, and they satisfy $D_{OS}=D_{LS}+D_{OL}$. $\triangle\alpha_{n}=\alpha-2n\pi$ is
the offset of the deflection angle, and $n$ is an integer that indicates the number of loops done by the photon around the black hole. Since the lensing effects are more significant when the objects are highly aligned, we will study the case which the angles $\beta$ and $\theta$ are small. We can find that the angular separation between the lens and the $n$th relativistic image is
\begin{equation}\label{theta}
\theta_{n}\simeq\theta^{0}_{n}+\frac{u_{ps}e_{n}
(\beta-\theta_{n}^{0})D_{OS}}{\bar{a}D_{LS}D_{OL}},
\end{equation}
with
\begin{equation}\label{theta1}
\theta_{n}^{0}=\frac{u_{ps}}{D_{OL}}(1+e_{n}),~~~~
e_{n}=e^{\frac{\bar{b}-2n\pi}{\bar{a}}},
\end{equation}
where $\theta_{n}^{0}$ is the image position corresponding to $\alpha=2n\pi$. As $n\rightarrow\infty$, we can find that $e_{n}\rightarrow0$ from Eq. (\ref{theta1}), which implies that the minimum impact parameter $u_{ps}$ and the asymptotic position of a set of images $\theta_{\infty}$ obey a simple form
\begin{equation}\label{theta2}
u_{ps}=D_{OL}\theta_{\infty}.
\end{equation}
Then, the magnification of the $n$th relativistic image is given by
\begin{equation}\label{magnification}
\mu_{n}=\left. \frac{1}{\frac{\beta}{\theta}\frac{\partial\beta}{\partial\theta}}\right|_{\theta^{0}_{n}}
=\frac{u_{ps}^{2}e_{n}(1+e_{n})D_{OS}}{\bar{a}\beta D^{2}_{OL}D_{LS}}.
\end{equation}
It is easy to find that the first relativistic image is the brightest, and the magnification decreases exponentially with $n$. Therefore, we only consider that the outermost and brightest image $\theta_{1}$ is resolved as a single image, and all the remaining ones are packed together at $\theta_{\infty}$ \cite{Bozza2,Bozza}. Thus, the angular image separation $s$ between the first image and the packed others, and the ratio $\mathcal{R}$ of the flux  from the first image to those from all other images can be expressed as
\begin{eqnarray}\label{s r}
 \nonumber   s&=&\theta_{1}-\theta_{\infty}=\theta_{\infty} e^{\frac{\bar{b}-2\pi}{\bar{a}}},\\
\mathcal{R}&=&\frac{\mu_{1}}{\Sigma^{\infty}_{n=2}\mu_{n}}=e^{\frac{2\pi}{\bar{a}}}.
\end{eqnarray}
These two formulas can be easily inverted to get
\begin{eqnarray}\label{ab2}
\nonumber \bar{a}&=&\frac{2\pi}{\log \mathcal{R}},\\
\bar{b}&=&\bar{a}\log\left(\frac{\mathcal{R}s}{\theta_{\infty}}\right).
\end{eqnarray}
For a given theoretical model, the strong deflection limit coefficients $\bar{a}$ and $\bar{b}$ and the minimum impact parameter $u_{ps}$ can be obtained; then these three observables $s$, $\theta_{\infty}$, and $\mathcal{R}$ can be calculated. On the other hand, comparing them with astronomical observations, it will allow us to determine the nature of the black hole stored in the lensing.

\begin{figure}[htbp]
\begin{center}
\includegraphics[scale=1]{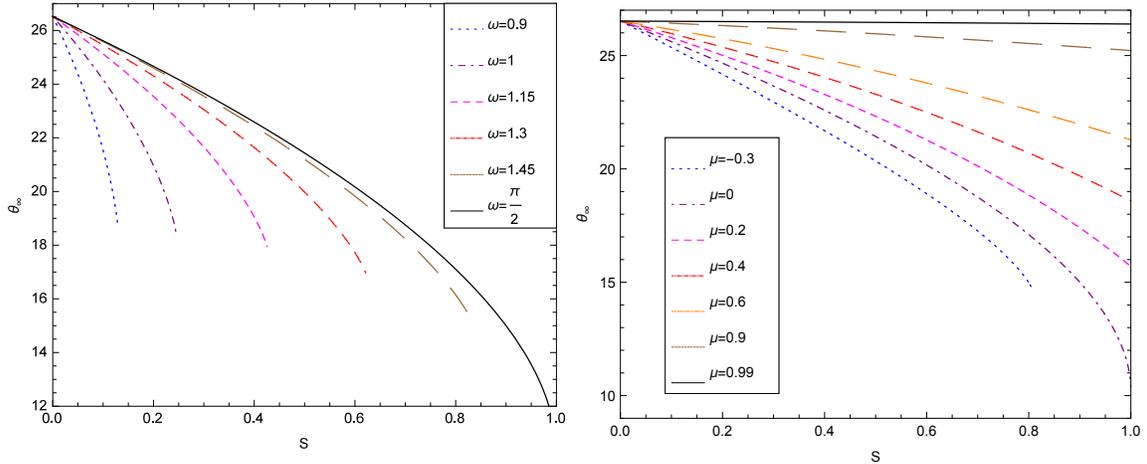}
\end{center}
\caption{Strong gravitational lensing by the Galactic center black hole. Variation of the angular image position $\theta_{\infty}$ which is expressed in \emph{$\mu$ arc seconds}  with scalar hair $S$ for $\mu=0$ (left) and $\omega=\frac{\pi}{2}$ (right).}
\label{th}
\end{figure}

\begin{figure}[htbp]
\begin{center}
\includegraphics[scale=1.02]{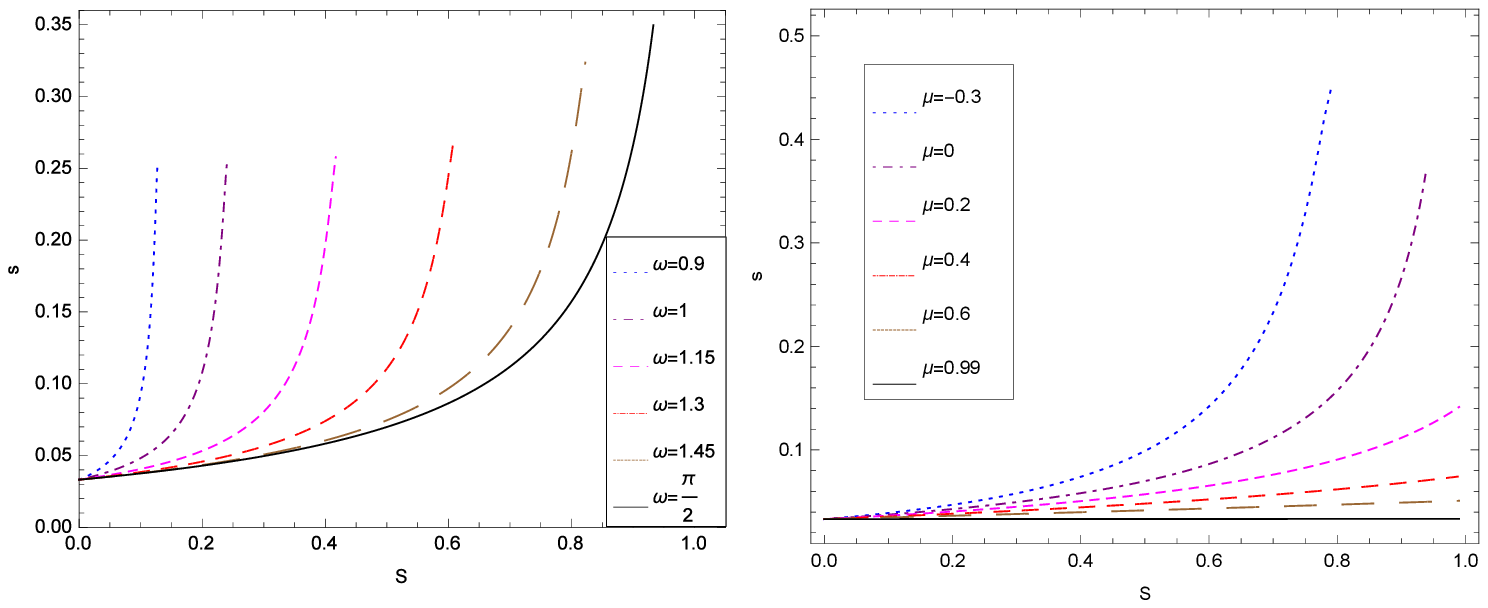}
\end{center}
\caption{Strong gravitational lensing by the Galactic center black hole. Variation of the angular image separation $s$  which is expressed in \emph{$\mu$ arc seconds} with scalar hair $S$ for $\mu=0$ (left) and $\omega=\frac{\pi}{2}$ (right).}
\label{s}
\end{figure}

\begin{figure}[htbp]
\begin{center}
\includegraphics[scale=1]{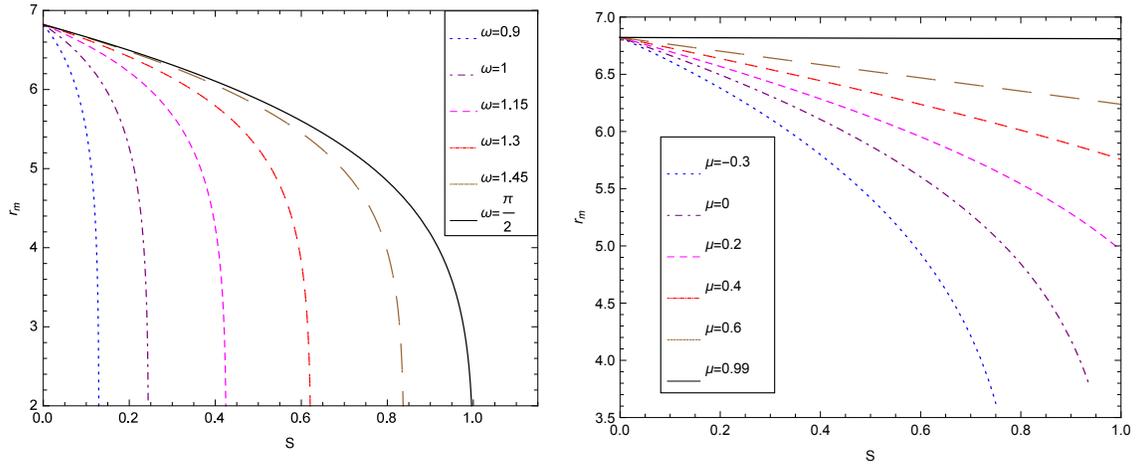}
\end{center}
\caption{Strong gravitational lensing by the Galactic center black hole. Variation of the relative magnitude $r_{m}$ with scalar hair $S$ for $\mu=0$ (left) and $\omega=\frac{\pi}{2}$ (right).}
\label{rm}
\end{figure}

To provide an example, let us consider the supermassive black hole in the Galactic center can be described by this solution. It has a mass $M=4.4\times10^{6}M_{\odot}$ \cite{Genzel}, and it is situated at a distance from the Earth $D_{OL}= 8.5$kpc; so the ratio of the mass to the distance is $\frac{M}{D_{OL}}\approx2.4734\times10^{-11}$. Hence, we can estimate the value of the coefficients and observables for strong gravitational lensing by combing with Eqs. (\ref{angle4}),  (\ref{theta2}), and (\ref{s r}). We present the numerical value for the angular image position $\theta_{\infty}$, the angular image separation $s$, and the relative magnitude $r_{m}$ (which is related to $\mathcal{R}$ by $r_{m}=2.5\log \mathcal{R}$) of the relativistic images in Figs. \ref{th}-\ref{rm}. We can find that the angular image position $\theta_{\infty}$ decreases with the increase of scalar hair $S$ for either $\mu=0$ case or $\omega=\frac{\pi}{2}$ case in Fig. \ref{th}. We also find that the $\theta_{\infty}$ grows with the increase of $\omega$ for fixed $S$ and $\mu$, and so does the change of $\theta_{\infty}$ with $\mu$ for fixed $S$ and $\omega$. Figures \ref{u} and  \ref{th} show that the changes in $\theta_{\infty}$ and $u_{ps}$ are the same, this is because $\theta_{\infty}$ and $u_{ps}$ satisfy the geometrical relationship of $u_{ps}=D_{OL}\theta_{\infty}$. Furthermore, we get from Figs. \ref{s} and \ref{rm} that the angular image separation $s$ increases, while the relative magnitude $r_{m}$ decreases with the increase of scalar hair $S$. It is interesting to find that for different $\omega$ and $\mu$, each line of $\theta_{\infty}$, $s$, and $r_{m}$ intersects at $S\rightarrow0$, which returns to the results of the standard Schwarzschild case for $\theta_{\infty}=26.5095 \mu$ arc sec, $s=0.0331 \mu$ arc sec, and $r_{m}=6.82$ \cite{Bozza}. There is another situation ($\omega=\frac{\pi}{2}$, $\mu\rightarrow1$) that can return to the results of the Schwarzschild case, they are plotted with black lines in the graph on the right of Figs. \ref{th}-\ref{rm}.

\section{Summary}

In this paper, we investigated strong gravitational lensing for the four dimensions charged black holes with scalar hair in the Einstein-Maxwell-Dilaton theory \cite{Fan}. We studied the effects of scalar hair on the event horizon $r_H$, the radius of the photon sphere $r_{ps}$, the strong deflection limit coefficient $\bar{a}$, the minimum impact parameter $u_{ps}$, the deflection angle $\alpha(\theta)$ and the main observables, such as the angular image position $\theta_{\infty}$, the angular image separation $s$, and the relative magnifications $r_m$ of relativistic images in strong gravitational lensing. In our usual perception, $r_{ps}$ is always greater than $r_H$ for an arbitrary black hole, Fig. \ref{rps00}, which showed the $r_H$ and $r_{ps}$ for the charged black holes with scalar hair, illustrated this view again. And Fig. \ref{rps00} also showed that the $r_H$ and $r_{ps}$ are both decrease with the increase of scalar hair.
We found from Figs. \ref{a} and  \ref{aa} that both the deflection angle $\alpha(\theta)$ and strong deflection limit coefficient $\bar{a}$ increase with the increase of scalar hair  for either $\mu=0$ or $\omega=\frac{\pi}{2}$, which means that the deflection angle of light ray is dominated by the logarithmic term in gravitational lensing.
We learned from Figs. \ref{u} and \ref{th} that the changes of the angular image position $\theta_{\infty}$ and the minimum impact parameter $u_{ps}$ are the same (both decrease with the increase of scalar hair for either $\mu=0$ or $\omega=\frac{\pi}{2}$) due to $\theta_{\infty}$ and $u_{ps}$ satisfy the geometrical relationship of $u_{ps}=D_{OL}\theta_{\infty}$. Moreover, we also found,  with the increase of  scalar hair, that the angular image separation $s$ increases, while the relative magnitude $r_{m}$ decreases from Figs. \ref{s} and \ref{rm}.
It should be pointed out that this black hole  can recover to the Schwarzschild black hole in two cases, one is $S\rightarrow0$ for arbitrary $\omega$ and $\mu$, another is $\omega=\frac{\pi}{2}$ and $\mu\rightarrow1$ for arbitrary scalar hair, and all quantities of strong gravitational lensing for the charged black holes with scalar hair can be reduced to those of the Schwarzschild spacetime--- i.e., $r_{H}=1$, $r_{ps}=1.5$, $\bar{a}=1$, $u_{ps}=2.598$, $\alpha(\theta)=6.28$, $\theta_{\infty}=26.5095\mu$ arc sec, $s=0.0331\mu$ arc sec, $r_{m}=6.82$.
This can be seen clearly in every figure, all the lines intersect at $S\rightarrow0$ and the black lines in the graph on the right side of figures, which stand for $\mu\rightarrow1$ with $\omega=\frac{\pi}{2}$, are basically straight lines.

\begin{acknowledgments}
{
{This work is supported by the  National Natural Science Foundation
of China under Grant Nos. 11475061;
the Hunan Provincial Innovation Foundation for Postgraduate (Grant No.CX2016B164).}}

\end{acknowledgments}

\end{document}